# Bound states in Bose-Einstein condensates with radially-periodic spin-orbit coupling

Chunyan Li,[1,2,*] Vladimir V. Konotop,[3] Boris A. Malomed,[4,5] and Yaroslav V. Kartashov[2]

[1]School of Physics, Xidian University, Xi'an, 710071, China
[2]Institute of Spectroscopy, Russian Academy of Sciences, 108840, Troitsk, Moscow, Russia
[3]Departamento de Física and Centro de Física Teórica e Computacional, Faculdade de Ciências, Universidade de Lisboa, Campo Grande, Ed. C8, Lisboa 1749-016, Portugal
[4]Department of Physical Electronics, School of Electrical Engineering, Faculty of Engineering, and Centre for Light-Matter Interaction, Tel Aviv University, 69978 Tel Aviv, Israel
[5]Instituto de Alta Investigación, Universidad de Tarapacá, Casilla 7D, Arica, Chile

We consider Bose-Einstein condensate (BEC) subject to the action of spin-orbit-coupling (SOC) periodically modulated in the radial direction. In contrast to the commonly known principle that periodic potentials do not create bound states, the binary BEC maintains multiple localized modes in the linear limit, with their chemical potential falling into spectral gaps of the (numerically found) radial band structure induced by the spatial modulation of the SOC. In the presence of the repulsive nonlinearity, the SOC modulation supports fundamental gap solitons of the semi-vortex types, as well as higher-order vortex gap solitons. The localization degree and stability of the gap solitons strongly depend on the location of their chemical potential in the gap. Stability intervals for vortex gap solitons in a broad range of the intrinsic vorticity, from $-2$ to $3$, are identified. Thus, the analysis reveals the previously unexplored mechanism of linear and nonlinear localization provided by the spatially periodic modulation of SOC, which may be extended to other settings, such as those for optical beams and polaritons. Unlike the commonly known quartets of eigenvalues for small perturbations, in the present system the instability is accounted for by shifted complex eigenvalue pairs.

## 1. Introduction

The significance of solitons as fundamental modes in a great variety of nonlinear media is commonly known [1,2]. Most studies in this field were focused on one-dimensional (1D) solitons, which have been observed in a huge number of experimental settings. A more challenging subject is the study of multidimensional solitons, the fundamental problem being that the ubiquitous cubic self-attraction, represented by the nonlinear Schrödinger (NLS) equation, gives rise to the critical and supercritical collapse of bright modes in 2D and 3D geometries, respectively [3-6]. The collapse destabilizes scalar and vector bright solitons (in particular, the critical collapse drives the slowly developing (sub-exponential) instability of the *Townes solitons* in the respective settings [3-6]), suggesting one to seek for physically relevant multidimensional settings which make it possible to stabilize them [7-9]. In particular, the observation of nearly stable 2D solitons of the Townes type in a binary Bose-Einstein condensate (BEC) was reported recently [10]. A still more challenging objective is stabilization of vortex solitons with bright ring-like shapes, i.e., multidimensional solitons with embedded vorticity, as they, unlike fundamental solitons, are vulnerable to the splitting instability, which may destroy the vortices even if the collapse is suppressed [9].

A stabilization mechanism which may be implemented in various physical contexts is provided by spatially periodic (lattice) potentials. In addition to the well-known realizations of lattices which support stable solitons in photonics [11-16], spatially periodic potentials induced by optical lattices (OLs) are widely used for the creation of various modes in BECs [17,18]. In particular, it was demonstrated that the square-shaped [19-21] and hexagonal [22] OLs stabilize 2D vortex solitons in the form of multi-peak patterns, with the vorticity represented by phase circulation along a contour connecting the peaks. Actually, the most natural setup for the stabilization of vortices is offered by circular OLs, as they conserve the angular momentum, which is linked to the vorticity [23-26]. In particular, a 2D OL with a radial structure can be induced in a BEC by a properly structured cylindric optical beam.

More recently, the use of spatially uniform [27-29] or modulated [30] spin-orbit coupling (SOC) between two components of a binary BEC was elaborated as a mechanism for the stabilization of 2D solitons carrying intrinsic vorticity. While in most cases SOC is implemented in the 1D form [31,32], it has been realized in 2D as well [33]. For this setting, it was predicted that SOC, which couples Gross-Pitaevskii equations (GPEs) for two components of the binary BEC, prevents the onset of the critical collapse and maintains *stable* 2D fundamental solitons, in the form of *semi-vortices* (SVs) [27,28], also known as half-vortex solitons [34]. The concept of half-vortices was formulated in the theory of quantum fluids [35], they were predicted [36] and observed [37,38] in exciton-polariton condensates. The respective solutions are characterized by the phase singularity and unitary vorticity in one component and no vorticity in the second component. Furthermore, the realization of SOC makes it possible to consider its version with the local strength subject to spatial modulation. In particular, solitons supported by spatially localized SOC were studied in 1D [39,40] and 2D [30,41] geometries. Thus, in Ref. [30] it was shown that an inhomogeneous SOC can sustain 2D linear localized modes, as well as ones in BEC with self-repulsion. The interplay of uniform SOC and a radially periodic potential was considered too [42].

In this work we aim to produce stable 2D modes in the binary BEC under the action of SOC periodically modulated in the radial direction. The first result is that 2D bound states are supported by the radial SOC lattice *without any nonlinearity*, while, as is well known, radial lattice potentials do not create bound (localized) states [25,27]. Then we demonstrate that, in the presence of the natural repulsive nonlinearity, the system supports vast families of stable gap solitons, which are both fundamental SVs and higher-order states, carrying vorticity in their both components, while the higher-order solitons are completely unstable in the case of the interplay of the spatially uniform SOC and self-attractive nonlinearity [27,28].

The rest of the paper is organized as follows. The model, based on the spinor GPE for the two-component mean-field wave function, is introduced in Section 2. Systematic numerical results demonstrating the existence of truly localized bound states supported by the radial SOC lattice in the purely linear version of the system are reported in Section 3 (recall this is principally impossible in any lattice potential possessing translational invariance). In Section 4, the analysis is extended for fundamental and higher-order gap solitons in the full nonlinear system, identifying their existence and stability domains. These solitons, which are bright modes in both components, originate from the bound states in the linear system.

## 2. The model: spinor Gross-Pitaevskii equations

The action of SOC and intrinsic nonlinearity with equal strengths of the intra- and inter-species repulsion on the binary BEC is modeled by GPEs for the two-component mean-field wavefunction $\Psi = (\psi_+, \psi_-)^T$. In the scaled spinor form, it is written as [43]

$$i\partial_t \Psi = \frac{1}{2}[-i\nabla + \mathbf{A}(r)]^2 \Psi + (\Psi^\dagger \Psi)\Psi, \quad (1)$$

where $\mathbf{r} = (x,y)$, $\nabla = (\partial_x, \partial_y)$, $\mathbf{A}(r) = \alpha(r)(-\boldsymbol{\sigma}_y, \boldsymbol{\sigma}_x)$, $\boldsymbol{\sigma}_{x,y}$ are the Pauli matrices, and

$$\alpha(r) = \alpha_0 \sin(\Omega r + \theta) \quad (2)$$

is the SOC strength periodically modulated along radial coordinate $r$ with amplitude $\alpha_0$, wavenumber $\Omega$, and phase shift $\theta$.

In the experiment, this setting can be realized by means of the technique developed for the creation of the effectively two-dimensional SOC in the condensate of $^{87}$Rb atoms [33], with a difference that the SOC-inducing laser beams will be shaped so as to form a radial lattice, instead of the square-shaped one used in the reported experiment. Relevant values of the lattice spacing and number of atoms in the condensate should be $2\pi/\Omega \sim$ a few $\mu$m and $N \sim 10^5$, with the atomic density $\sim 10^{14}$ cm$^{-3}$.

Writing Eq. (1) in polar coordinates $(r, \phi)$, see Eq. (A1) in Appendix, we conclude that axisymmetric stationary states are looked for as (see also Ref. [27])

$$\psi_+(r,\phi,t) = u_+(r)e^{-i\mu t + i(S-1)\phi}, \quad \psi_-(r,\phi,t) = u_-(r)e^{-i\mu t + iS\phi}, \quad (3)$$

with topological charge (winding number) $S$ and chemical potential $\mu$. Real radial functions $u_\pm(r)$ introduced in Eq. (3) satisfy the following equations:

$$\begin{aligned}
\mu u_+ &= -(1/2)\big[\partial_r^2 + r^{-1}\partial_r - r^{-2}(S-1)^2\big]u_+ + \alpha^2 u_+ - \\
&\quad (1/2)\alpha'_r u_- - \alpha\big(\partial_r + r^{-1}S\big)u_- + \big(u_+^2 + u_-^2\big)u_+, \\
\mu u_- &= -(1/2)\big[\partial_r^2 + r^{-1}\partial_r - r^{-2}S^2\big]u_- + \alpha^2 u_- + \\
&\quad (1/2)\alpha'_r u_+ + \alpha\big[\partial_r - r^{-1}(S-1)\big]u_+ + \big(u_+^2 + u_-^2\big)u_-.
\end{aligned} \quad (4)$$

where $\alpha'_r \equiv \partial_r \alpha$ [recall $\alpha(r)$ is defined as per Eq. (2)].

We aim to construct localized states vanishing at $r \to \infty$. Their existence domain with respect to chemical potential $\mu$ is determined by the SOC-modulation landscape, defined by Eq. (2). One can see this by analyzing the asymptotic solution at $r \to \infty$, where the nonlinear terms and linear terms $\sim r^{-1}$, $r^{-2}$ may be neglected, reducing Eq. (4) to 1D linearized coupled equations with $r$-periodic coefficients. Being formally considered on the entire real axis, their solutions are Bloch waves $u_{\pm,\kappa} = w_{\pm,\kappa}(r)e^{i\kappa r}$ with a periodic function $w_{\pm,\kappa}(r + 2\pi/\Omega) = w_{\pm,\kappa}(r)$, and quasi-momentum $\kappa$ taken in the first Brillouin zone, $|\kappa| < \pi/\Omega$. In accordance with the Bloch theory, eigenvalues $\mu = \mu(\kappa)$ form bands separated by gaps, that are shown by gray and white areas, respectively, in Fig. 1(a) as functions of the SOC strength $\alpha_0$. Localized states in such SOC landscapes may exist only for values of $\mu$ belonging to spectral gaps of the linear radial Bloch spectrum.

## 3. Bound (localized) states in the linear system with the radial SOC lattice

The radially periodic SOC lattice creates localized states even in the absence of the nonlinearity. Eigenvalues of such states, obtained numerically by solving linearized version of Eq. (4):

$$\begin{aligned}
\mu u_+ &= -(1/2)\big[\partial_r^2 + r^{-1}\partial_r - r^{-2}(S-1)^2\big]u_+ + \alpha^2 u_+ - \\
&\quad (1/2)\alpha'_r u_- - \alpha\big(\partial_r + r^{-1}S\big)u_-, \\
\mu u_- &= -(1/2)\big[\partial_r^2 + r^{-1}\partial_r - r^{-2}S^2\big]u_- + \alpha^2 u_- + \\
&\quad (1/2)\alpha'_r u_+ + \alpha\big[\partial_r - r^{-1}(S-1)\big]u_+,
\end{aligned} \quad (5)$$

where we now keep all linear terms $\sim r^{-1}$, $r^{-2}$, to get solutions on the entire axis $r \geq 0$ (in contrast to the asymptotic analysis at $r \to \infty$ described above, that makes it possible to identify the borders of the allowed bands), are shown in Figs. 1(a) and 1(f) for two different modulation profiles (2), viz., $\alpha(r) = \alpha_0 \sin(\Omega r)$ and $\alpha(r) = \alpha_0 \cos(\Omega r)$ by lines with circles. Similar to the common situation for gap solitons in systems with self-repulsive nonlinearities [17,18], branches of eigenvalues of the bound states in the linear system are entirely embedded in the first finite gap of the spectrum plotted in Figs. 1(a,f) (i.e. such modes may be maintained solely in the gap). For eigenvalues $\mu$ that are placed deeply inside the gap, the corresponding eigenmodes are well-localized ones [see Figs. 1(b) and 1(d)], while they gradually delocalize when $\mu$ approaches one of the gap edges [see Fig. 1(c)].

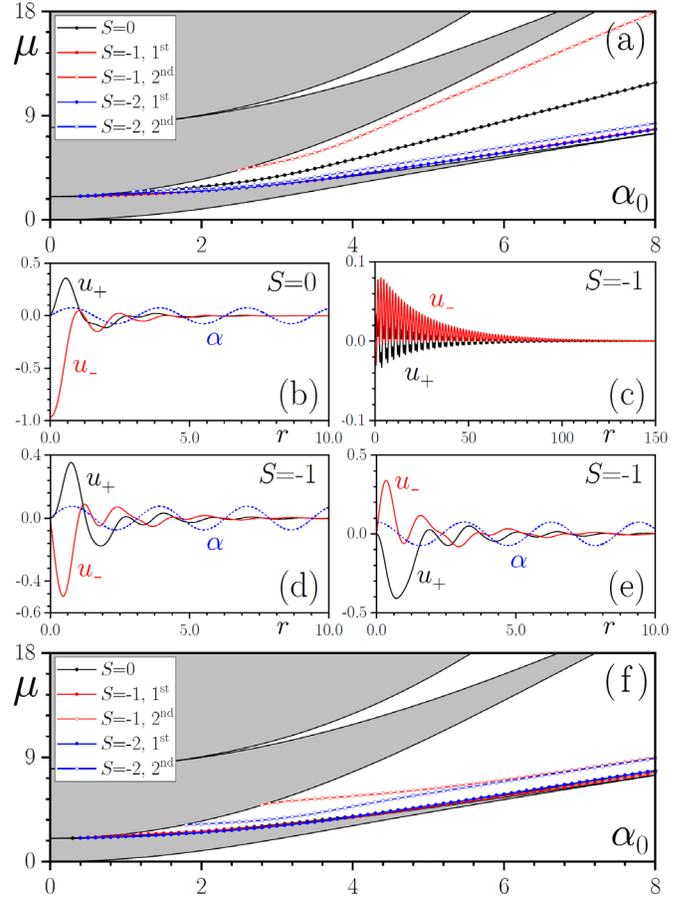

Fig. 1. Eigenvalues of bound states, produced by the numerical solution of the *linearized system* of Eq. (5), vs. the SOC amplitude $\alpha_0$ in the first finite radial gap for the sine (a) and cosine (f) SOC profiles, see Eq. (2). Radial Bloch bands and gaps occupy gray and white areas, respectively. Black curves with dots designate eigenvalues of the single branch of bound states of the linear system with $S=0$, while red and blue curves designate two branches of the bound states with winding numbers $S=-1$ and $S=-2$, respectively, which are defined according to Eq. (3). Panel (a) demonstrates that all branches of the bound (localized) states are embedded in the first finite gap. Examples of the bound states with $S=0$ (b) and $S=-1$, belonging to the first (c) and second (d) branches, that are supported by the sinusoidal SOC modulation profile with $\alpha_0 = 3$ in Eq. (2) (SOC profile is shown by the dashed blue lines). Linear eigenvalues of corresponding states are $\mu \approx 3.741$ (b), $\mu \approx 4.955$ (c), and $\mu \approx 3.052$ (d). Panel (e) shows the eigenmode from the second branch with $S=-1$, $\mu \approx 5.101$, supported by the cosine SOC profile (2) with $\alpha_0 = 3$. The same value of the modulation amplitude is used in all figures below. In all panels, the radial wavenumber is fixed as $\Omega = 2$ while amplitude $\alpha_0$ is considered as variable parameter that can be always done using scaling transformations for Eq. (1).

Due to the time-reversal symmetry of Eq. (5), if $(u_+, u_-)$ is a solution with vorticity $S \leq 0$, then $(u_-, -u_+)$ is a solution with vorticity $1-S$ corresponding to the same $\mu$ possessing the same properties [27]. Therefore, we consider only the states with $S = 0, -1, -2, \ldots$, while states with topological charges $S = 1, 2, 3, \ldots$ can be easily constructed from known solutions with $S = 0, -1, -2, \ldots$ using the above symmetry properties. We have found a single branch of the linear localized states for $S=0$ and two branches for each $S<0$ in the first finite gap of the spectrum, see Figs. 1(a,f). Among these two branches, one with smaller $\mu$ (shown by lines with solid circles these figures) is located close to the bottom gap's edge, hence the corresponding eigenmodes are loosely bound, as seen in Fig. 1(c). The second branch with larger $\mu$, shown by lines with open circles in Figs. 1(a,f), usually emerges near the top gap's edge and delves deeper into the gap with the increase of $\alpha_0$, leading to shrinkage of the corresponding localized eigenmodes. Note that eigenvalues of the localized modes are different for the cosine [Fig. 1(f)] and sine [Fig. 1(a)] SOC modulation profiles in Eq. (2), both of them supporting loosely and tightly bound states, cf. Figs. 1(d) and 1(e). Below we focus on $\alpha(r) = \alpha_0 \sin(\Omega r)$, the results for the cosine profile being quite similar.

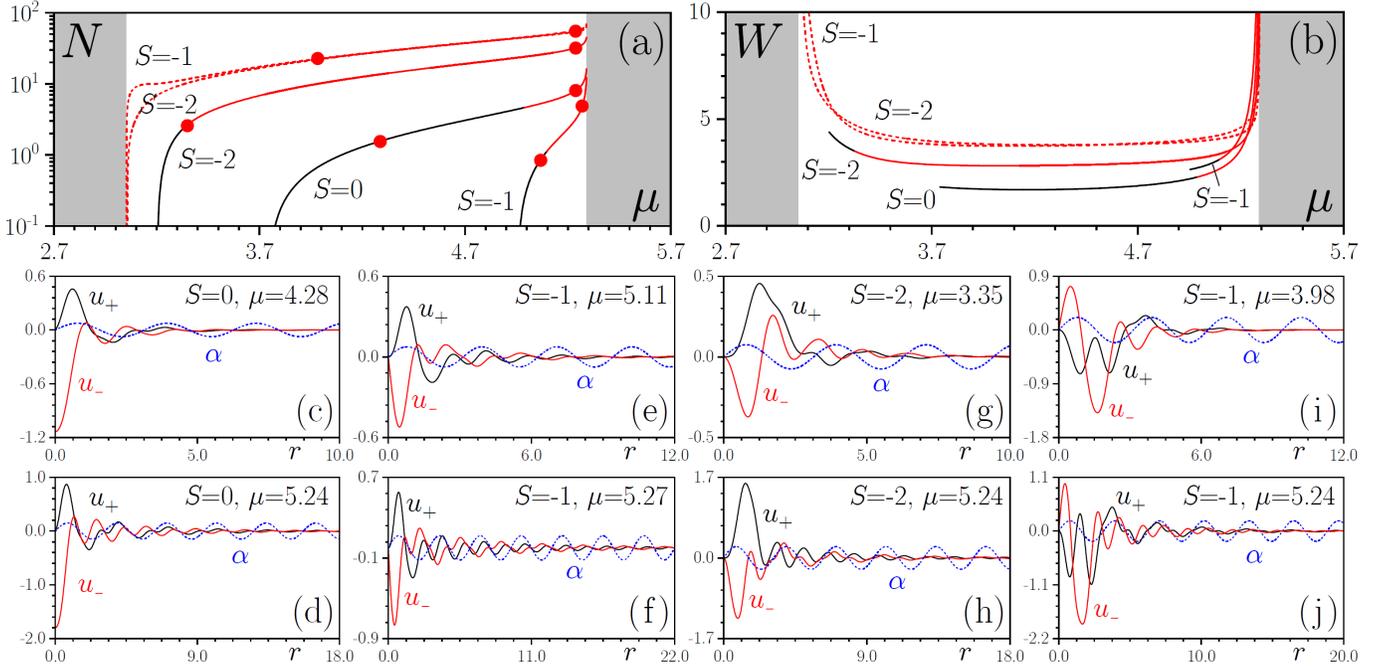

Fig. 2. Norm $N$ (a) and width $W$ (b) vs. chemical potential $\mu$ for gap solitons with different vorticities (winding numbers) $S$, defined as per Eq. (3). The soliton families bifurcating, under the action of the self-repulsive nonlinearity, from the first and second branches of the linear bound states are shown by dashed and solid lines, respectively. Gray areas designate spectral bands, in which gap solitons do not exist. Black and red segments designate stable and unstable gap-soliton families, respectively. (c)-(j) Examples of gap solitons with different vorticities, corresponding to red dots in (a). In panels (e)-(h) and (i,j) the gap solitons bifurcate from the linear bound state belonging to the second and first branches, respectively.

## 4. Families of fundamental and higher-order gap-vortex solitons

In the presence of the repulsive nonlinearity, gap solitons bifurcate from the linear modes considered above. To produce such modes, we solved Eq. (4) with nonlinear terms numerically by dint of the Newton method. The gap solitons feature radial oscillations of the wavefunction, indicating that they emerge from Bragg reflections in the radial direction, produced by the SOC modulation. The gap-soliton norm $N = \int \mathbf{\Psi}^\dagger \mathbf{\Psi}\, d\mathbf{r}$ is plotted in Fig. 2(a) as a function of $\mu$ for different vorticities $S$ and a characteristic SOC amplitude in Eq. (2), $\alpha_0 = 3$. The $N(\mu)$ dependences for the solitons emanating from the first and second branches of the linear bound states are shown by dashed and solid lines, respectively. For all gap-soliton families, the norm is naturally vanishing when $\mu$ approaches the respective linear eigenvalue, and monotonously increases towards the top edge of the gap. In the latter limit, gap solitons tend to delocalize, developing extended radial tails. This is demonstrated by the dependence of the integral radial width, $W = N^{-1} \int \mathbf{\Psi}^\dagger \mathbf{\Psi} r d\mathbf{r}$, on $\mu$ in Fig. 2(b).

Because gap solitons belonging to the first family bifurcate from the linear branch close to the bottom of the gap, they are very wide: their width in the vicinity of the bottom of the spectrum is dictated by the width of the corresponding linear mode localized due to SOC. The growth of $\mu$ corresponds to the growth of the number of atoms, i.e., of the repulsive nonlinearity, which becomes a dominating factor responsible for the localization. Notice that near the bottom of the first gap the effective dispersion is negative, as determined by the effective periodicity introduced by SOC. It is thus natural that repulsive nonlinearity initially stimulates localization of such very wide states close to the bottom of the gap. In contrast, gap solitons belonging to the second family remain well localized at the point where they emerge from corresponding linear branch. Examples of gap-soliton profiles corresponding to red dots in Fig. 2(a) are displayed in 2(c-j).

The simplest state, corresponding to $S=0$ in Eq. (3) and shown in Fig. 2(c) is the SV carrying vorticity in the $u_+$ component only. The profile of soliton displayed in Fig. 2(c) is close to profile of linear mode from Fig. 1(b), since its norm $N \approx 1.538$ is not too high and it can be considered as nonlinear deformation of linear state from which it bifurcates. However, close to the top edge of the gap the amplitude of this state substantially increases and it acquires long oscillating tail, as shown in Fig. 2(d). Profiles of more complex gap solitons with $S=-1$ and $S=-2$ in Eq. (3), i.e., both components carrying vorticities, that bifurcate from the second family of the localized linear eigenmodes, are presented in Figs. 2(e,f) and 2(g,h), respectively. These solitons are tightly and loosely localized in the depth of the gap and close to its top edge, respectively. Figures 2(i,j) present examples of gap solitons bifurcating from the first branch of the linear bound states. Such states feature a tightly-localized shape close to the center of the gap. Note that, even though the increase of SOC amplitude $\alpha_0$ in Eq. (2) results in tighter localization, it does not necessarily lead to expansion of stability domains, in contrast to the known results for gap solitons in usual OLs, where deeper periodic potentials imply wider areas if the stability in the respective parameter spaced [17].

The central result of this work is that in the general case the radially periodic SOC sustains stable gap solitons with various values of winding number $S$, defined as per Eq. (3), even if both components carry nonzero vorticity. To show this, we have performed the linear stability analysis of the obtained solitons (see Appendix for the details). We search for perturbed solutions as

$$\psi_+(r,\phi,t) = [u_+(r) + p_+(r)e^{\lambda t + iM\phi} + q_+^*(r)e^{\lambda^* t - iM\phi}]e^{-i\mu t + i(S-1)\phi},$$
$$\psi_-(r,\phi,t) = [u_-(r) + p_-(r)e^{\lambda t + iM\phi} + q_-^*(r)e^{\lambda^* t - iM\phi}]e^{-i\mu t + iS\phi},$$
(6)

where $\lambda = \lambda_{\mathrm{re}} + i\lambda_{\mathrm{im}}$ is the perturbation growth rate, $p_\pm, q_\pm$ are profiles of perturbation eigenmodes, $M$ is the integer azimuthal perturbation index (independent of $S$), and asterisks stand for complex conjugation. The substitution of ansatz (5) in Eq. (1) and linearization yields the linear eigenvalue problem for $\lambda$, that was solved numerically. The solution is stable if $\lambda_{\mathrm{re}} \leq 0$ for all $M$. The so produced (in)stability predictions were checked by direct simulations of the perturbed evolution of the solitons in the framework of Eq. (1). The results of the stability analysis are summarized in Figs. 2(a,b), where black and red lines designate stable and unstable gap solitons, respectively. The soliton family with $S=0$ features the largest stability interval $\mu \in [3.741, 4.990]$, as it loses stability only near the top edge of the gap, which is the generic trend known for gap solitons [17]. Remarkably, the soliton families with both $S=-1$ and $S=-2$, bifurcating from the second branch of the localized linear eigenmodes, as defined above, also have their stability domains for our parameters [thus, $S=-1$ state is stable for $\mu \in [4.955, 5.098]$, while $S=-2$ state is stable for $\mu \in [3.203, 3.330]$]. These stability intervals are adjacent to the points at which such vortex-carrying gap solitons bifurcate from the linear bound states.

The stability of the vortex solitons in this system is possible because the repulsive nonlinearity tends to suppress the azimuthal instability, to which vortex modes are vulnerable in the case of self-attraction [7]. Indeed, the linear stability analysis shows that instabilities for the gap solitons with $S=-1$ and $S=-2$, that set in near the top gap's edge, are typically very weak. Figure 3 shows a typical spectrum of stability eigenvalues in the $(\lambda_{\mathrm{re}}, \lambda_{\mathrm{im}})$ plane for dominant azimuthal perturbations with $M=1$ [Fig. 3(a)] and $M=-1$ [Fig. 3(b)] for the gap soliton emerging from the second linear mode branch with $S=-1$ at $\mu=5.2$. Small nonzero real parts $\lambda_{\mathrm{re}}$ of complex eigenvalues in Figs. 3(a,b) indicate the trend of this vortex- gap soliton to decay in the course of long evolution, as observed in Fig. 4(d). An unusual feature is that, instead of simultaneous appearance of a quartet of eigenvalues $\{\lambda, -\lambda, \lambda^*, -\lambda^*\}$, which is the usual scenario [3,4,6], only a pair $\{\lambda, -\lambda^*\}$ exists for both $M=1$ and $M=-1$. Moreover, for $M=1$ only perturbation modes $p_\pm(r)$ from Eq. (6) are well localized [see Figs. 3(c,e)], while modes $q_\pm(r)$ are weakly localized, see Figs. 3(d,f) [the situation is reversed for $M=-1$, when $q_\pm(r)$ are well localized instead], indicating that a large radial domain is necessary for accurate computation of the perturbation spectrum in this case. These peculiarities are explained by the fact that the SOC terms in Eqs. (1) reduce the symmetry of the system. Note that, for the perturbation eigenmode with $M=0$, one still observes the usual situation with a quartet of the stability eigenvalues. Finally, all solitons bifurcating from the *first* families of the linear bound states are unstable, as seen in Fig. 2(a) [corresponding instability intervals are $\mu \in [3.052, 5.291]$ for $S=-1$ state and $\mu \in [3.056, 5.291]$ for $S=-2$ state].

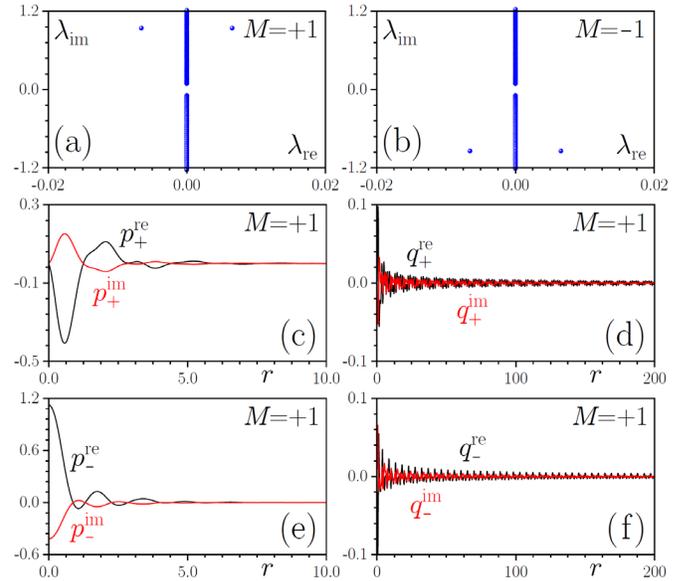

Fig. 3. The eigenvalue spectrum for the azimuthal perturbations with indices $M=1$ (a) and $M=-1$ (b), for the soliton with $\mu=5.2$ and $S=-1$. (c)-(f) Profiles of the perturbation eigenmodes [defined according to Eq. (6)] corresponding to the unstable eigenvalue with $M=1$ and $\lambda_{\mathrm{re}} > 0$.

To verify the predictions of the stability analysis, in Figs. 4(a),(c), and (e) we display examples of long stable evolution of initially weakly perturbed gap solitons bifurcating from the second branch of the linear bound states, with vorticities $S=0$, $-1$, and $-2$. The evolution was modelled using the Crank–Nicolson algorithm with the five-point discretization scheme for spatial derivatives (along each spatial coordinate). We used $dt=0.001$ as the time step, and $dx=dy=0.05$ as the spatial ones. In Fig. 4 the absolute value and phase are shown only for the first component of the gap-vortex solitons, $|\psi_+|$ and $\varphi_+ = \arg(\psi_+)$. Stable solitons maintain their shapes in the course of very long evolution. Note that an unstable soliton of the SV type, with $S=0$ and $\mu=5.2$ [Fig. 4(b)] spontaneously transforms, through emission of radiation, into a similar SV state, but with a smaller norm. On the other hand, the unstable vortex with $S=-1$ and $\mu=5.2$, shown in Fig. 4(d), exhibits the onset of azimuthal instability, as a result of which it transforms into an SV soliton with $S=0$, but much later. The stable evolution of the higher-order vortex soliton with $S=-2$ and $\mu=3.3$ is shown in Fig. 4(e), while unstable solutions of this type with larger $\mu$ split in two azimuthal fragments under the action of the dominant perturbation with perturbation numbers $M=\pm 2$, see Eq. (6).

We have also investigated the structure and stability of GSs in the framework of Eq. (1) with the attractive nonlinearity, which corresponds to the opposite sign in front of the cubic term, and can be implemented in the experiment by means of the well-known Feshbach-resonance technique [44]. In that case, all the higher-order solitons, with $S \leq -1$, resemble the "excited states" considered in Ref. [27], and, similar to them, they are unstable.

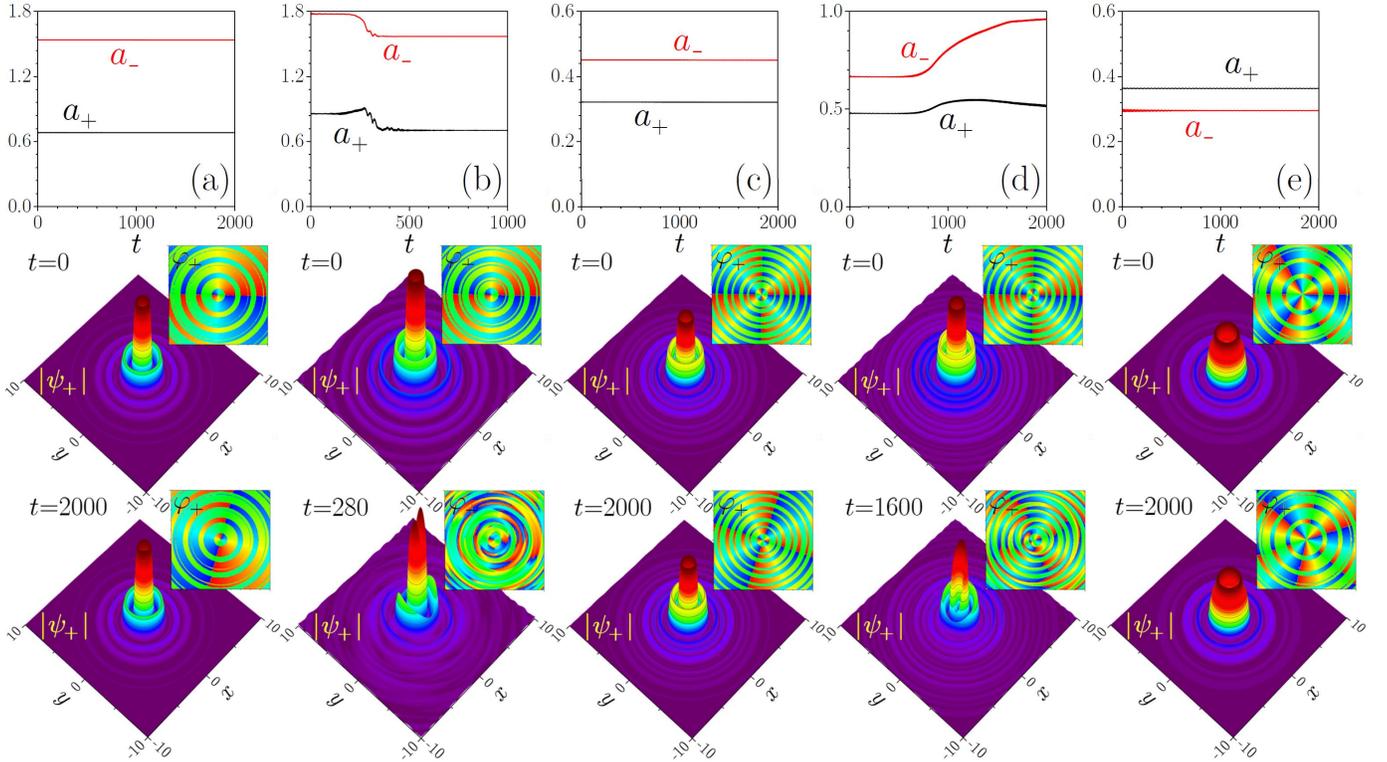

Fig. 4. Examples of the stable evolution of the gap solitons with (a) $S=0$, $\mu=4.8$, (c) $S=-1$, $\mu=5.07$, and (e) $S=-2$, $\mu=3.3$. Also displayed is the evolution of unstable solitons with (b) $S=0$, $\mu=5.2$ and (d) $S=-1$, $\mu=5.2$. Peak amplitudes $a_\pm = \max|\psi_\pm|$ are shown, as functions of time, in the top row. Other panels display the comparison of initial and final profiles of the absolute values and phases of component $\psi_+$.

## 5. Conclusion

We have introduced the 2D system which offers a new possibility for the creation of bound (localized) states, in both linear and nonlinear settings. The system is based on the GPE for the spinor wave function of the binary BEC whose components are coupled by the spin-orbit interaction subject to periodic modulation in the radial direction. We have found that, even in the absence of the nonlinearity, the system supports truly localized ground states, in the form of semi-vortices, and higher-order bound states with embedded additional vorticity, which is impossible in quantum mechanics with periodic potentials. Chemical potentials of these bound state belong to the first finite spectral gap of the linearized system. Self-repulsive nonlinearity transforms the linear bound states into gap solitons, whose stability and radial width are strongly affected by the proximity of their chemical potential to edges of the finite spectral gap. The stability regions have been found for the gap solitons with many values of the intrinsic vorticity (winding number), from $-2$ to $3$. An unusual feature revealed by the analysis is that unstable eigenvalues appear not in usual quartets, but as pairs shifted in the complex plane.

It may be relevant to extend the analysis for two-component solitons of the bright-dark type [45]. A challenging possibility is to construct solitons in the 3D settings, where uniform SOC supports metastable SVs [46]. Another relevant direction is to introduce radial SOC lattice for polaritons in semiconductor microcavities, where the vorticity difference between the components is 2, rather than 1 in the binary BEC [47,48], and for SOC-emulating optical dual-core couplers [49].

**Acknowledgements:** We appreciate valuable discussions with E. Y. Sherman. This research is funded by the research project FFUU-2021-0003 of the Institute of Spectroscopy of the Russian Academy of Sciences. C. Li acknowledges a grant of National Natural Science Foundation of China (NSFC) (11805145) and China Scholarship Council (CSC) (202006965016). The work of B.A.M is supported, in part, by the Israel Science Foundation through grant No. 1695/22. The work of V.V.K. is supported by the Portuguese Foundation for Science and Technology (FCT) under Contracts PTDC/FIS-OUT/3882/2020 and UIDB/00618/2020.

*chunyanli@xidian.edu.cn

**Data availability**

Data will be made available upon reasonable request.

**Declaration of Competing Interest**

Authors declare that they have no conflict of interest.

**Credit authorship contribution statement**

All authors contribute greatly to the work.

**Appendix: Stability analysis and linear eigenvalue problem**

The linear stability analysis of the stationary gap-soliton solutions is based on the system of Gross-Pitaevskii equations [Eq. (1) in the main text], written here in polar coordinates $(r,\phi)$:

$$i\frac{\partial}{\partial t}\begin{pmatrix}\psi_+\\\psi_-\end{pmatrix}=-\frac{1}{2}\begin{pmatrix}\partial_r^2\psi_+ + r^{-1}\partial_r\psi_+ + r^{-2}\partial_\phi^2\psi_+\\\partial_r^2\psi_- + r^{-1}\partial_r\psi_- + r^{-2}\partial_\phi^2\psi_-\end{pmatrix}+\alpha^2(r)\begin{pmatrix}\psi_+\\\psi_-\end{pmatrix}+$$
$$\begin{pmatrix}|\psi_+|^2+|\psi_-|^2\\|\psi_-|^2+|\psi_+|^2\end{pmatrix}\begin{pmatrix}\psi_+\\\psi_-\end{pmatrix}+\frac{1}{2}\begin{pmatrix}-\psi_-e^{-i\phi}\partial_r\alpha(r)\\\psi_+e^{+i\phi}\partial_r\alpha(r)\end{pmatrix}+$$
$$\alpha(r)\begin{pmatrix}-e^{-i\phi}[\partial_r\psi_- - (i/r)\partial_\phi\psi_-]\\e^{+i\phi}[\partial_r\psi_+ + (i/r)\partial_\phi\psi_+]\end{pmatrix} \quad (A1)$$

We search for perturbed solutions in the form given by Eq. (6) in the main text. In this work, we consider azimuthal perturbation indices $M=0,\pm 1,\pm 2,\ldots\pm 5$, as it is unlikely that the perturbations with higher indices may lead to instability. The linearization of Eq. (A1) around stationary solution $u_\pm(r)$ yields the eigenvalue problem for $\lambda$:

$$\begin{aligned}i\lambda p_+ &= -\mu p_+ - (1/2)[\partial_r^2 + r^{-1}\partial_r - r^{-2}(S+M-1)^2]p_+ + \alpha^2 p_+ - (1/2)\alpha_r p_- - \alpha[\partial_r + r^{-1}(S+M)]p_-\\&\quad +(2u_+^2 p_+ + u_+^2 q_+) + (u_-^2 p_+ + u_+ u_- p_- + u_+ u_- q_-),\\i\lambda q_+ &= +\mu q_+ + (1/2)[\partial_r^2 + r^{-1}\partial_r - r^{-2}(S-M-1)^2]q_+ - \alpha^2 q_+ + (1/2)\alpha_r q_- + \alpha[\partial_r + r^{-1}(S-M)]q_-\\&\quad -(u_+^2 p_+ + 2u_+^2 q_+) - (u_-^2 q_+ + u_+ u_- p_- + u_+ u_- q_-),\\i\lambda p_- &= -\mu p_- - (1/2)[\partial_r^2 + r^{-1}\partial_r - r^{-2}(S+M)^2]p_- + \alpha^2 p_- + (1/2)\partial_r\alpha p_+ + \alpha[\partial_r - r^{-1}(S+M-1)]p_+\\&\quad +(2u_-^2 p_- + u_-^2 q_-) + (u_+^2 p_- + u_+ u_- p_+ + u_+ u_- q_+),\\i\lambda q_- &= +\mu q_- + (1/2)[\partial_r^2 + r^{-1}\partial_r - r^{-2}(S-M)^2]q_- - \alpha^2 q_- - (1/2)\partial_r\alpha q_+ - \alpha[\partial_r - r^{-1}(S-M-1)]q_+\\&\quad -(u_-^2 p_- + 2u_-^2 q_-) - (u_+^2 q_- + u_+ u_- p_+ + u_+ u_- q_+).\end{aligned} \quad (A2)$$

Equation (A2) was solved numerically by means of a standard eigenvalue solver, producing the spectrum of $\lambda$ for different azimuthal perturbation indices $M$. The appearance of $\lambda_{\rm re}>0$ indicates the instability of the soliton. Notice that the unusual structure of instability spectrum presented in Fig. 3 for states with nonzero vorticity (with only pair of unstable eigenvalues) follows from symmetry of linearized Eqs. (A2) dictated by SOC terms. Thus, for a given sign of $M$ and for nonzero topological charge $S$ of soliton these equations can be rewritten to acquire identical form upon transformation $\lambda\Rightarrow -\lambda^*$, but it is not possible to rewrite them in identical form for example for $\lambda\Rightarrow\lambda^*$ transformation.